# OPEN SOURCE CRM SYSTEMS FOR SMES

Marco Tereso[1] and Jorge Bernardino[1]

[1] Polytechnic of Coimbra – ISEC, Coimbra, Portugal
a21190968@alunos.isec.pt, jorge@isec.pt

**ABSTRACT**

*Customer Relationship Management (CRM) systems are very common in large companies. However, CRM systems are not very common in Small and Medium Enterprises (SMEs). Most SMEs do not implement CRM systems due to several reasons, such as lack of knowledge about CRM or lack of financial resources to implement CRM systems. SMEs have to start implementing Information Systems (IS) technology into their business operations in order to improve business values and gain more competitive advantage over rivals. CRM system has the potential to help improve the business value and competitive capabilities of SMEs. Given the high fixed costs of normal activity of companies, we intend to promote free and viable solutions for small and medium businesses. In this paper, we explain the reasons why SMEs do not implement CRM system and the benefits of using open source CRM system in SMEs. We also describe the functionalities of top open source CRM systems, examining the applicability of these tools in fitting the needs of SMEs.*

**KEYWORDS**

*CRM, Information Systems, Open Source, SMEs.*

## 1. INTRODUCTION

As companies face increasing competition, wider customer choice, and the challenges of doing e-Business in the 21st century, many have chosen to implement CRM solutions in response to their strategic imperative, and to improve the sales and marketing effectiveness and efficiency. A Customer Relationship Management (CRM) system is a software system designed to empower a company to maximize profits by reducing costs and increasing revenue; to increase competitive advantage by streamlining operations; and to achieve business goals. Most companies have been collecting information about their customers and trying to use this information to better understand and predict what customers might want next. Therefore, CRM is both information-based and technology based, and is about building customer loyalty by putting customers at the center of what a company does. CRM also applies to how customers experience a company, not just how a company looks at its customers [1].

To survive in the global market, focusing on the customer is becoming a key factor for big and small companies. It is known that it takes up to five times more money to acquire a new customer than to get an existing customer to make a new purchase [2]. Therefore, customer retention is in particular important to Small and Medium Enterprises (SMEs) because of their limited resources. A second aspect of CRM is that knowing the customer and his/her problems allows a company to acquire new customers more easily and facilitates targeted cross-selling [3].

The CRM is a set of processes and methodologies that has as main objective to bring together a wide range of information about customers and their consumption behaviour. This type of information allows companies to define groups of users, making possible some sales strategies and campaigns and promotions targeted at different types of customers.

Open source software for CRM represents a great opportunity for SMEs as they can significantly impact their competiveness. Given the high fixed costs of normal activity of companies, we intend to promote open source solutions for small businesses. In this paper, we present viable solutions, describing the functionalities of top CRM tools, according to recent publications [4, 5]. We will describe each of these tools to assess their capabilities, their strengths and the aspects that differ between them, exploring the applicability of each of those tools in fitting the needs of SMEs.

This paper is structured as follows. In Section 2 we give a brief introduction about Information Systems (IS) and CRM, and presents the reasons why SMEs do not implement CRM systems. In Section 3 we list the advantages of open source software. In Section 4 we present CRM open source systems, with an overview of its features and Section 5 describes related works. Finally, in Section 6 we present our conclusions and point out future work.

## 2. INFORMATION SYSTEMS AND CRM

Given the competitive market, companies must adopt strategies that allow them to remain active in the market and become increasingly competitive. The collection of data resulting from business activity is extremely important. This process allows data to work in different areas, in order to make business activity most profitable.

Data is one of the most important and valuable assets for companies and organizations. The grouped data when generating information is a very important role to reduce uncertainty. The recognition of the importance of information by business managers led to the development of information systems. Information systems may have different objectives, taking into account the type of information we have to process and the desirable results.

The focus of our work is in IS (Information Systems) area, primarily CRM (Customer Relationship Management) systems. IS are systems that present a set of elements that capture, store, process and communicate information. There are several types of IS, with each type designed for a particular purpose.

In this section we intend to specify the importance of collecting data, that later on turns information in knowledge. We will also explain the importance to implement information systems into companies and organizations.

We present CRM systems in greater detail, analyzing their characteristics and importance to companies and organizations to disseminate information regarding the advantages of their use.

### 2.1. Why SMEs do not implement CRM system

In this competitive market, the client must be regarded as the fundamental entity of an organization [6]. The importance of satisfying the customer must be seen as a priority. Given the importance that clients represent to the enterprises economy, especially in SMEs, organizations must adopt strategies that allow them to define the profile of its customers. It is in this context that the CRM software has great importance in collecting and recording data.

The data should be regarded as one the most important assets of an organization, that start in information and will be transformed in knowledge. The organizations have in their data, the possibility of its future growth. Later using a Decision Support System (DSS) is possible to identify strengths and areas for improvement in order to optimize business processes and customer relations. Given all these advantages we must ask why companies do not implement CRM systems?

The current economic crisis leads companies to adopt strategies in order to fight against the difficulties. Known the swollen costs of normal business activity, it becomes difficult to SMEs to make extra investments. Lack of knowledge by SMEs of the existence of free solutions (open source) is one factor that determines why these tools are not acquired and implemented in the organizations. Our main goal is to disseminate to SMEs some viable solutions to implement a free CRM system to improve their business processes.

## 3. ADVANTAGES OF USING OPEN SOURCE SOFTWARE

Before going further, it is essential answer an important question, that is, which is the advantage of using open source software?

Regarding the costs, the purchase of open source software is free and the hardware associated to these systems has lower costs. This is justified, taking into account that those which are looking for open source software are not available to purchase commercial software, and also usually don't have financial resources to invest in more sophisticated equipment.

The open source software has a good quality, despite being developed by community. Nowadays the releases and the stable versions of software usually suffer an extended inspection and are exposed to various tests of software quality before being delivered.

Often the commercial software lacks some essential functionalities for organizations. The fact of commercial software does not provide the source code, not allows adding extra features. The open source software provides its source code, enabling organizations to develop new characteristics and integrating them into their applications.

The integration of other complementary software, is an additional possibility. The availability of the source code allows the integration of different tools. For example, it is possible the integration of open source databases, open source CRM systems, open source business intelligence tools, etc.

Simple license management, without license fees, without expired licenses date, etc. is a clear advantage of using open source software. Open source software usually has a support community, documentation, discussion forums and constant updates of its versions, updates, identify and resolve software glitches and bugs.

In summary, the main advantages of using open source software by companies are: free license, low cost maintenance, support community; the software can be shared and used for various purposes, access to source code and permission to study and amendment, ability to adapt to the real needs of each organization, constant updating of versions through the contribution of the community that supports it, possibility to try the software without any cost and ease of access to the specific open source repositories to download [25, 26].

For SMEs the main benefit of using open source CRM tools is the ability to adapt the software to business needs, with opportunity to add new modules developed accordingly to the company requirements. The possibility to access the code allows the company to benefit of the initially features available but also to configure the application taking into account their needs and environment.

Today it is very important that companies make cost containment. The SMEs has here good solutions that allow the evolution of business and reduce the expenses.

## 4. OPEN SOURCE CRM TOOLS

In an era of financial crisis and expense reduction it is important that companies keep pace with changing technologies and markets, adopting strategies for tracking trends, but which are not expensive. SME's need low cost CRM solutions that can easily adapt to their business model and IT structure, instead of having to adapt their business model and IT structure to the CRM software. In this section our main objective is to disclose open free solutions viable for implementation by SMEs.

In this particular case we rely on some recent sources of information, namely [4] and [5]. We will describe open source CRM systems to assess their capabilities, their strengths and the aspects that differ between them and exploring the applicability of each of those tools in fitting the needs of SMEs. The open source CRM systems that we analyze are: CiviCRM [7], OpenCRX [8], OpenTaps [9], SugarCRM [10], Vtiger [11] and Free CRM [12].

### 4.1. CiviCRM

CiviCRM [7] is an open source tool, based on relationship management solutions. This tool is web-based, multi-language, and intended to needs of advocacy, non-profit and non-governmental groups. CiviCRM allows recording data for the various constituents of an organization as volunteers, activists, donors, customers, employees, suppliers, etc.

The main strengths of the tool are: store information about individuals, organizations and households; the community of development work regularly in the growth and extension of system; ease of integration with other websites; CiviCRM integrates directly into the popular open source Content Management Systems – Drupal and Joomla; save the access history of each user, using the log record; CiviCRM was designed for use worldwide, allowing the many localized formats and supporting multi language, such as, French, Spanish, German, Dutch, Portuguese, etc.; licensed under the GNU AGPL allow a changes in code, add new features, etc. [7].

The advantages of the web-based applications are the possibility of linking several people at the same time in different locations, accessibility within and outside the corporate infrastructure, making communication between people practical and accessible.

CiviCRM has optional components that provide a better service of support to its users. The CiviCRM components are: CiviContribute – component that allows managing online donations and contributions to an organization; CiviCase – to manage interactions between people and organizations; CiviEvent – allows the creation and listing of paid and free events, allows to create events through the creation of Web pages specifically for this purpose; CiviMail – e-mail service that lets redirect to the lists and email contacts and various custom reports; CiviMember – to on-line registration and revalidation of the user account; CiviReport – module tool that allows to create reports on data that are intended to illustrate [7].

When associate with a CMS (Content Management System) CiviCRM allows a content management relatively to customers data. With the implementation of the CMS tool the visitors have the possibility to signing up for events, requesting email updates and donating money.

CiviCRM is a web tool that can be installed on a Web server or an internal server. The requirements for installing CiviCRM are open source tools, namely: Apache Server 2.0 (or higher), PHP 5.2.1 (or though PHP 5.3 is only supported from version 3.2 of CiviCRM), MySQL 5.1.x (or higher), Drupal 6.x ou 7.x/Joomla 1.5.x ou 1.6.x, Server Cronjobs and require 128 Mb of memory [7].

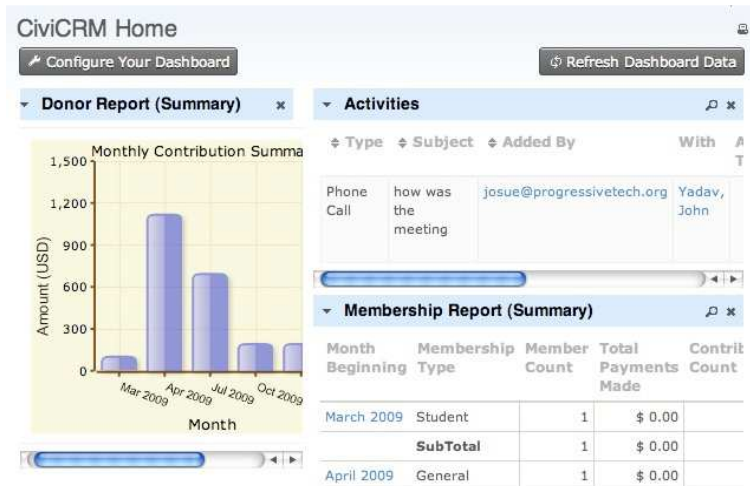

Figure 1. CiviCRM ambient work [20]

The Figure 1 illustrates the application screenshot with several elements, charts, tables, information, etc. The left side of the figure is visible a chart of donor report, the right of the figure in top is visible a panel of activity log and in bottom a membership report summary.

## 4.2. OpenCRX

In 2003 with the need to create a free CRM tool, the Swiss company Crixp Corporation supported by the Swiss company Omex AG, developed the OpenCRX tool.

OpenCRX is a true open source tool, which has enterprise-class with features like role-based security, system-wide/pervasive audit-trail, platform independence, virtually unlimited scalability, and much more. OpenCRX besides the code also provides UML models and the full javadoc, that allow to the users more knowledge of the system, it is great for programmers [8].

OpenCRX is a tool developed in Java, which can run on any operating system by simply installing a JVM (Java Virtual Machine). This application lets the user to change the language.

The main characteristics of OpenCRX are [8]: Open source - an open source tool that allows to adapt real needs. The possibility of re-structure their UML architecture, makes it easier to amend it; Account Management - allows for a detailed consultation to contacts, activities, products and states of account; Monitoring of trade, show all possible approaches and prospects to customers; Management of products and their prices - manage the lists of products and their prices, enables the conversion of prices for different currencies. For the sales process, OpenCRX has the particularity to enable the integration of an ERP system for management of sales processes; Activity management - essential to the creation of groups and activities to control the number of hours wasted in solving problems; Administration - the security aspects are taken into account too, the tool allows some control extended as part of their modules; Integration with e-mail POP / IMAP / SMTP coming from different servers and services of e-mail; Integration with MS Office and Open Office suite tools; Time management - coordination of working hours and the reconciliation between the various tasks across the enterprise with customers, it is possible using the functionality CalDAV of the OpenCRX; Diversity of data support - the tool supports data different data management systems, such as, MySQL, Oracle, MS SQL Server, PostgreSQL and DB2.

OpenCRX includes business intelligence tools such as BIRT and the construction of reports is possible. The OpenCRX installation requires: JDK, JEE Server, Apache ANT and a SGBD.

According to [8] OpenCRX is a tool that requires more learning time compared with other similar open source CRM tools. The fact that it is not a tool based on LAMP (Linux, Apache, MySQL, PHP) makes it more difficult to install compared with solutions of this type. The ANT (is a project of Apache Software Foundation, allows building automation software) technology becomes more complex to implement.

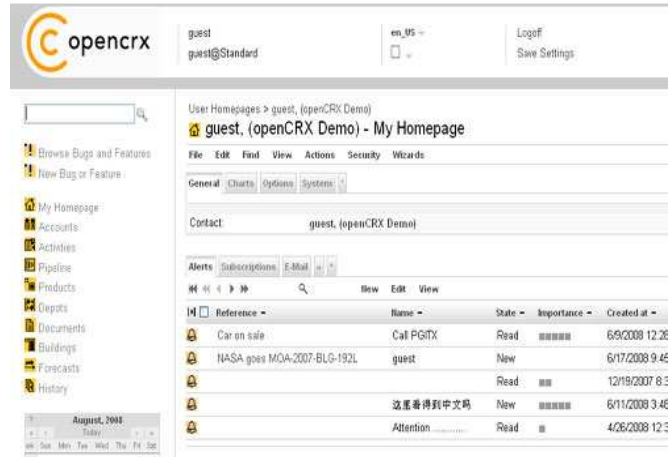

Figure 2. OpenCRX ambient work [21]

In Figure 2 is possible to visualize an application screenshot of OpenCRX. In the left it is a tool bar, in the right is visible the work ambient composed by tab options and content tables.

### 4.3. OpenTaps

OpenTaps [9] is an open source tool of CRM + ERP (Enterprise Resource Planning). OpenTaps was developed thinking in small and medium enterprises and had as its starting point the Apache Open for Business. OpenTaps was developed in Java language (JEE), which offers scalability of implementation.

OpenTaps has the following features [15]: eCommerce, Poin-of-sales, inventory, warehouse, order, customers management and general ledger. OpenTaps allows interconnection with business intelligence tools (iReport) and mobility integration with Microsoft Outlook, Google Calendar and mobile phones.

OpenTaps is many used by organizations that are linked to activities such: industrial machinery manufacturers, online good retailers, online content distributors, telecommunications companies, independent software vendors and hosted services providers. For these industries, OpenTaps is used for order management, customer service purchasing, production planning, inventory management, and manufacturing to shipping and accounting [9].

We enumerate some key features of the tool OpenTaps based on [15] and [16]: OpenTaps has a web based user interface; internationalization - multi language; integrate marketing and sales - customers services, warehouse, supply chain, online and physical stores, and accounting; marketing - support stores, catalogs, categories and products; promotion store - online promotion store; cross-sells (are suggested complementary products) and up-sells (update to sell any product or service) for products; price rules for customers or group-specific pricing; integration online and Point of Sales (POS); real-time accounting reports; manage the manufacturing process; plan and automate purchasing and manufacturing.

OpenTaps has also some additional modules namely: project management; human resources; fixed asset management; Outlook and mobile phone integration with Funambol; integration with

eBay, Google base and Google check out; content management; Point of Sales (POS); integration eCommerce; Business Intelligence (integration with Pentaho and JasperReports).

The OpenTaps can be applied in systems with the following characteristics: operating systems - Windows, Mac OS X, Linux and Solaris; SGBDs - Oracle, Microsoft SQL Server, MySQL and PostgreSQL; Google Web Toolkit (GWT), allows an use more easy of the OpenTaps.

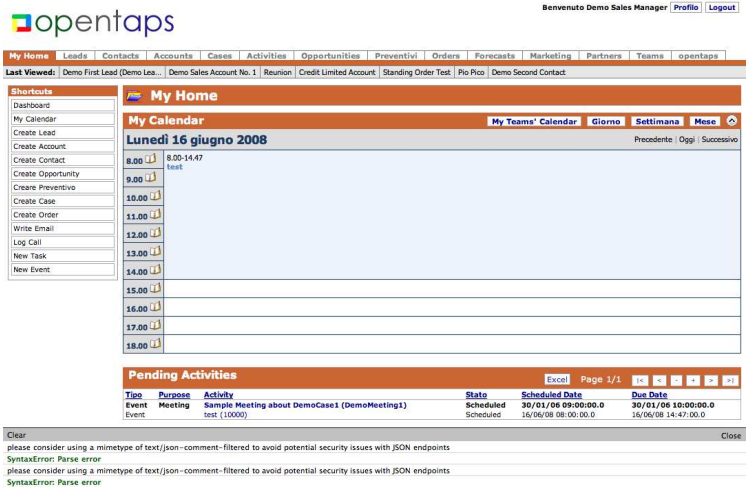

Figure 3. OpenTaps screenshot [22]

In Figure 3 is illustrated the work ambient of OpenTaps tool. In the left of the figure is visible a menu options, in top is visible some tabs of options. In the right of the figure is visible the user 'home' and in the bottom a table of pending activities.

### 4.4. SugarCRM

SugarCRM [10] is a software company based in California. SugarCRM includes some features, such as: sales-force automation, marketing campaigns, customer support, collaboration and reporting.

SugarCRM is one of the premier CRM solutions, which today has over 4 million of users. The open source version is developed in PHP, this project is growing rapidly and has an active developer community [30].

The main features of SugarCRM tool are [10]: scalability of pages to application; contacts management; relationship accounts; marketing campaigns; tasks and activities management; documental management; Web portal; occurrences management; processing emails.

The requirements to support the application are: operating systems - Linux, Windows or Unix; PHP 5.2.x (or higher); MySQL 4.1 (or higher) or MS SQL Server; Apache 1.3 (or higher), IIS 6 (or higher) or NGINX 1.x (or higher, with php-fpm); HD with 100MG; Around 64M of memory by user.

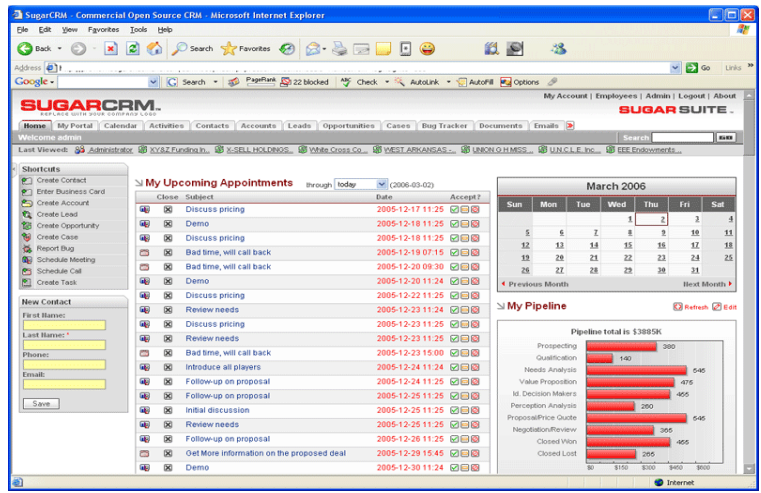

Figure 4. SugarCRM work ambient [23]

The Figure 4 illustrates the work ambient of SugarCRM. On the left in the figure is visible a menu options, in top exists some tabs, on the right is illustrate a calendar and a chart, on centre is visible a table with upcoming appointments of user. In comparison with other tools, SugarCRM present a big amount of icons and tabs.

### 4.5. Vtiger

Vtiger [11] is an open source business-oriented solution, built on platforms LAMP/WAMP (Linux/Windows, Apache, MySQL and PHP).

Vtiger has been developed by the same development community of SugarCRM, with the intention of developing an open source CRM solution and providing similar functionality to SugarCRM and Salesforce.com.

Vtiger CRM has 15 modules: covering Marketing, Sales, Support, Inventory, and project Management functions. Vtiger is available in more than 15 different languages[11].

The main features of Vtiger are: sales automation; control support; Marketing automation; inventory management; support for multiple database systems; security control; product customization; integrated calendar; integration with email service; add-ons (plug-in for Outlook, MS Office, Thunderbird, customers portal and web forms).

Vtiger software needs the following requirements: operating system - Linux or Windows; Apache 2.0.40 or higher; MySQL 5.1.x or higher; PHP 5.0x or higher; Firefox or Internet Explorer; minimum of 200 MB free disk space; minimum of 256MB RAM.

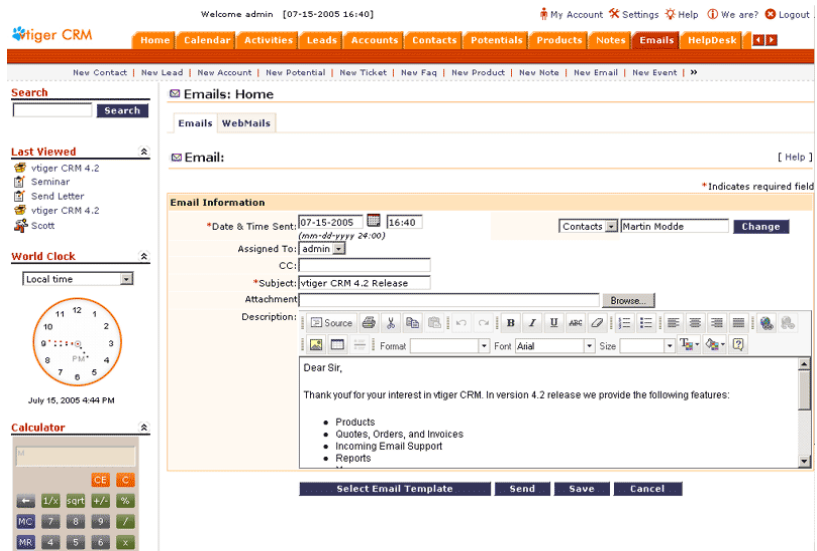

Figure 5. VtigerCRM screenshot [24]

The Figure 5 illustrating the appellative ambient of VtigerCRM tool. On the left in figure is visible a browser search, a list of last viewed, a time zone setting and a calculator. In top of figure is visible a set of option tabs. In centre of figure is visible the properties of email service.

### 4.6. Free CRM

Free CRM [12] is an open source CRM software, which is a Web based solution for customer relationship management and sales force automation. This is a software for up to 50 users that doesn´t require to download any software. In the open source version, Free CRM is somewhat limited with regard to storage capacity and security [25].

Free CRM is great for contact and lead tracking, sales and contact management, sales pipeline management and forecasting, customer services and business management [27].

The main features of FreeCRM are: contact and lead tracking; sales and pipeline management; support ticket & service management; advanced security & uptime; superior technical support; tasks; reports; calendar.

FreeCRM has some specifics available functionalities [12]:

- Calendar – schedule all of meetings and calls using the scheduler; track team member's availability and schedule; set multi-user alerts that send notification emails to all members.

- Company – track calls, history, trouble tickets, and sales activities; introduce workflow and automation to sales campaigns, vendors and prospects; bridge contact management data with tasks, deals, calendar events, and more.

- Contact – import contact lists from any spreadsheet file; synchronize the new contacts and related information in real time; empower full sales potential with an interface that allows to get down to business with contacts database.

- Deal – create a complete view of your sales process; analyze the progress any given campaign, cycle or employee is doing; quantify your sales success over time.

- Task – get the tasks priority; create custom tasks types specific for business; assign tasks to team members or employees; give workflow and status of the tasks; delegate workload across the team.

- Case – customize support issues with our custom fields; assign issues to one or multiple team members; track the cases so no customer issue falls through the cracks.

- Call – create call campaigns; automate calls; organize follow-up.

- Email – integration with the service of POP email.

- Doc – management documents.

- Form – give customer a quality feedback form; give your staff questionnaires and scripts; build custom surveys for sales, marketing, and customer satisfaction.

- Report – generate charts and graphs of company performance; generate hard numbers based on calls campaigns, email campaigns, tasks, deals, and much more; choose from over 70 pre-defined reports.

The requirements systems, necessary to run the application are [28]: Internet Explorer or Netscape.

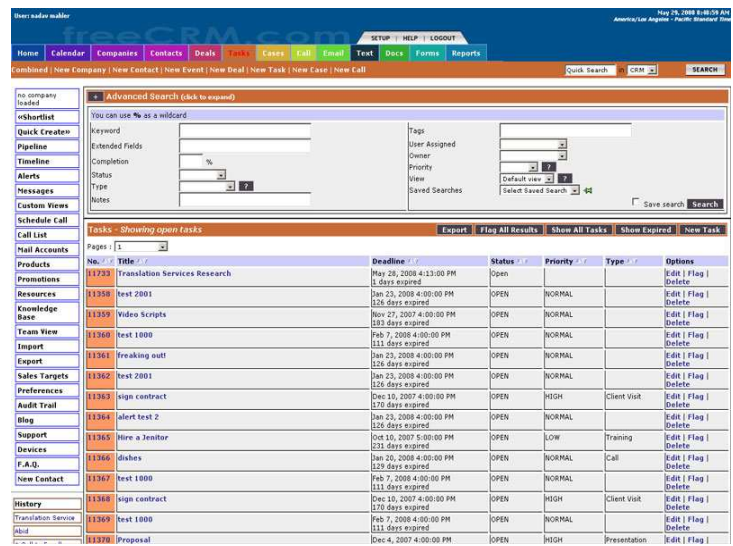

Figure 6. FreeCRM screenshot [12]

In Figure 6 is visible the ambient of work of the FreeCRM tool. In the top of figure is visible the option tabs, on the left is possible view the options menu, in centre of figure is visible the list of tasks.

## 5. RELATED WORK

According to a publication of Global Alliance of SMEs [31], SMEs represent over 98% of companies worldwide; based on this percentage we have focused our research into small firms with lower investment capacity. CRM systems are a necessity for companies and organizations to improve their competitiveness. However, there are not many papers focusing into open source CRM tools. For example [32] and [33], evaluate and present a review of the top 10 CRM open source solutions based on utility, business models and developer communities. In [32] they show a comparison between the top 10 CRM tools, containing a features table of each tool, helping to decide which is the best tool. In a white paper published by CRM Outsiders [34], an industry blog, and sponsored by SugarCRM is given a description about advantages and disadvantages of the use the CRM systems by SMEs. In [35] is presented the evaluation about the costs between open source and proprietary solutions. The choice of CRM tool is very important, such as, the knowledge of the importance its use, for this purpose, references [36] and [37] may be useful. In [38] is described an application for evaluating CRM open source systems. And [39] presents a description about a comparison about assessment methodologies between free and open source software. However, companies sponsor most of these works and there are not many research papers. In our work we presented an independent evaluation of open source CRM tools.

## 6. CONCLUSIONS AND FUTURE WORK

Nowadays, the contact lists, the customers profile information and a basic knowledge about customers have a great importance for the enterprises and organizations. Data is the most important asset in any organization, from which it is possible to extract information and transform it into knowledge.

The main objective of this work was to promote some open source CRM solutions for small enterprises and organizations. The adoption of open source CRM tools has many advantages for SMEs, as the possibility of really trying the system, reduction of vendor lock-in, low licence cost and possibility of in-depth personalization. Taking into account the importance of data for enterprises and organizations, is essential the acquisition these tools by enterprises and organizations.

All the described tools, CiviCRM, OpenCRX, OpenTaps, SugarCRM, Vtiger and Free CRM, have the particularity of being open source, not entail any licence costs. Each CRM solutions analyzed here has its particularities, their advantages when compared with others, different features, different difficulties of using and differences in system requirement.

When choosing of the CRM tool to adopt, is important that the final customer evaluate the tool taking into account the requirements of the organization and evaluate the features of each tool. The easiness of adapting the tool is one of the aspects to consider.

It is often difficult to evaluate which the best solution CRM to implement in organizations, taking into account the best solution is also conditioned by the specific needs of each company. Taking into account the lack of related works, and how we do not implement in practice each of the tools, would not be fair to define which/what the best of these solutions. The sources [4] and [5] elect the same tools as the best open source CRM today. This is a good idea for a future work, examine in practice, what the top tree best CRM open source systems.

Overall, the CRM tools described here, have the potentiality for help small enterprises and organizations in their day to day, contributing for contacts management of customers, tasks management, save the customers data, and define customer profile. Among all the features the most important is the fact to be open source and without licence costs. From our evaluation, we conclude that we should not elect any of the tools as the winner. All analyzed tools have capacities that can be adopted by companies and organizations and all have quality to serve the needs and requirements of SMEs.

We can conclude the world of open source tools has good solutions to be implemented in small and medium enterprises. This type of systems also allows the organizations to modernize and optimize their business processes at low costs.

**Authors**

Short Biography

Jorge Bernardino is Coordinator Professor of the Department of Systems and Computer Engineering at the ISEC (Instituto Superior de Engenharia de Coimbra) of Polytechnic of Coimbra, Portugal. He was President of ISEC from 2005–2010 and President of Scientific Council during 2003–2005. He received his PhD Degree in Computer Engineering from the Computer Engineering Department of the University of Coimbra in 2002. He is member of Software and Systems Engineering (SSE) group of Centre for Informatics and Systems of the University of Coimbra (CISUC) research centre. His main research areas are Data Warehousing, Business Intelligence, database knowledge management, e-business and e-learning.

Marco Tereso studied Computer Science - Information Technology and Multimedia at ESTGOH (Escola Superior de Tecnologia e Gestão de Oliveira do Hospital) and attained his graduation in 2009. After completion of graduation joined the ISEC (Instituto Superior de Engenharia de Coimbra) where attends the Master of Informatic and Systems - Software Development. At this moment is to finish the master's thesis on open source business intelligence tools under the orientation of Prof. Jorge Bernardino. His main research areas are Business Intelligence and open source systems.